\documentstyle[12pt]{article}
\begin{document}
\begin{titlepage}
\title{
INSTABILITY OF SPACE-TIME DUE TO EUCLIDEAN WORMHOLES
}

\author{
   V.A.Rubakov and O.Yu. Shvedov  \\
{\small{\em Institute for Nuclear Research of the Russian
Academy of Sciences,  }}\\ {\small{\em 60-th October Anniversary Prospect
7a, Moscow 117312, Russia
}}\\ {\small and}\\ 
{\small {\em Sub-department of Quantum Statistics and Field Theory,}}\\
{\small{\em Department of Physics, Moscow State University }}\\
{\small{\em Vorobievy gory, Moscow 119899, Russia}} }

\end{titlepage}
\maketitle

\begin{flushright}
gr-qc/9608065
\end{flushright}

\begin{center}
{\small{\bf Abstract}}
\end{center}

{\small
The problem of topology 
change transitions in quantum gravity is discussed.
We argue that the contribution of the Giddings-Strominger wormhole
to the Euclidean path integral is pure
imaginary. This  is checked by two techniques: by the functional
integral approach and by the analysis of the Wheeler-De Witt equation.
We present also a simple quantum mechanical model which shares many 
features of the system consisting of parent and baby universes. In
this simple model, we show that quantum coherence is completely lost
and obtain the equation for the effective density matrix of the
''parent universe''.
}

\vspace{2cm}

This talk concerns the contribution of wormhole configurations, fig.1,
into the Euclidean functional integral for the forward ''flat space 
$\rightarrow$ flat space'' amplitude in quantum gravity. In many models,
there exist classical Euclidean wormhole solutions, which may be considered
as saddle points of this integral. The question is whether their 
contributions are real or imaginary, i.e. whether wormholes are analogous
to instantons or to bounces in quantum field theory, and what could be
their interpretation in the latter case. 

The simplest wormhole solution \cite{GiddingsStrominger} 
emerges in the theory of gravity
interacting with  massless three-form field (axion), or, equivalently
\cite{KLee}, of the theory of massless scalar field $\theta(x)$ minimally
coupled to gravity. The wormhole solution is $O(4)$ -- symmetric, 
\begin{equation}
ds^2=N^2(\tau)d\tau^2 + R^2(\tau) d\Omega_3^2,
\label{2*}
\end{equation}
where
\begin{equation}
\dot{\theta}=\frac{iq}{2\pi^2R^3\kappa^{1/2}}, \qquad
\dot{R}^2=1-\frac{q^2}{24\pi^4R^4}, \qquad N=1.
\label{2**}
\end{equation}
 Here $\tau$ is the coordinate running into the wormhole,
 $d\Omega_3^2$ is  metrics of the unit 3-sphere,
 $\kappa$ is the gravitational constant, and $q$ is an arbitrary parameter
 of the solution, which has  meaning of the global charge flowing into
 the wormhole. The values of the field $\theta$ are purely imaginary, 
 so that the wormhole contributes to the amplitude not in coordinate, but
 in momentum representation, since the momentum $p(x)$ conjugate to 
 $\theta(x)$ is real \cite{KLee,ColemanLee,AbbottWise}.

 The analysis 
 of $O(4)$ -- symmetric fluctuations has been recently performed
in ref.\cite{RShv}. 
It has been found that there exists a negative mode about
the wormhole solution. It has been argued, therefore, that the wormhole
contribution to the forward amplitude is imaginary. It was suggested in
ref.\cite{RShv} that the latter fact means instability of flat space-time
with respect to spontaneous emission of baby universes which in turn implies
the loss of quantum coherence in the large universe, as originally suggested
in refs.\cite{Hawking,LRTNPB}.
 Further support of this point of view comes from
the semiclassical analysis of the Wheeler-DeWitt equation for baby universe
that branches off \cite{ShvedovITEP}. The latter analysis reveals the role
of the boundary conditions for the wave function of the baby universe
that has to be imposed at singularity, $R=0$.

There are a few subtleties concerning the diagram of fig.1. The purpose of 
this talk is to discuss some of these subtleties in turn. 

{\bf 1}. For arbitrary positions of the wormhole ends in flat space-time,
the values of the field $\theta$ evaluated along the flat space and along
the wormhole do not match each other. In other words, the field $\theta$ 
has to be multivalued. 

The resolution of this subtlety is in fact well-known \cite{ColemanLee}.
Let us consider only those wormholes that carry fixed global charge $q$.
This is accomplished by inserting the
$\delta$-function:
$$
1=\int dq \delta(\int_S p({\bf x}) d{\bf x} -q).
$$
into the Euclidean functional integral, where $S$ is the 3-sphere shown in 
fig.1, and $p({\bf x})$, as before, is  momentum conjugate to
$\theta$. 
Representing this 
$\delta$-function as $\delta(\xi)=(2\pi)^{-1} \int d\lambda
e^{i\lambda\xi}$, one finds that the saddle-point equation for $\theta$ is
$$
\dot{\theta}= \lambda \delta(\tau-\tau_S) + 
\mbox{\rm smooth   function};
$$
this means that $\theta$ may indeed have the 
discontinuity on the surface $S$.

{\bf 2.} The wormhole ends can be placed at any two points $x$ and $y$ 
in flat space-time. At large $|x-y|$ the wormhole action is independent
of $|x-y|$. Therefore, the wormhole contribution is proportional to
$(VT)^2$ rather than usual $VT$, where $V$ and $T$ are normalization
spatial volume  and time, respectively.

This infrared peculiarity has to do with infinite degeneracy of vacuum of 
the field $\theta$ in infinite flat space: there exists a state with zero
energy for any value of the global charge $q$. To see the infrared nature 
of the factor $(VT)^2$, consider finite volume case (fig.2). One can 
show that at large time separation, $|x^0-y^0| \gg V^{1/3}$, the wormhole
action has the following dependence on $|x^0-y^0|$:
$$
S \sim \frac{q^2|y^0-x^0|}{2V} 
$$
Hence in finite volume, the integral over the positions of the wormhole ends
(''almost zero modes''), $\int dxdy e^{-S/\kappa}$, is proportional to $T$.
We conclude that in the theory with infrared cut-off, the dependence of the
wormhole amplitude on normalization time is exactly as it should be.

{\bf 3.} The integration over the $O(4)$ -- symmetric fluctuations about the
wormhole solution, as it stands, is ill defined because of the wrong sign
of the action for conformal factor in quantum gravity. To cure this problem,
the Gibbons-Hawking-Perry rotation \cite{GHP} has been performed in 
ref. \cite{RShv}. Even though GHP prescription works nicely in many cases,
additional arguments confirming the conclusion that the wormhole amplitude
of fig.1 is imaginary, are of importance.

One of these arguments comes from the observation that at large separation 
between the wormhole ends the diagram of fig.1 can be cut as shown in fig.3.
Since the negative mode found in ref.\cite{RShv} is $O(4)$ -- symmetric,
it sufficies to evaluate the latter diagram in the minisuperspace
approximation, leaving $R(\tau)$, $N(\tau)$, $\theta(\tau)$ and $p(\tau)$
as the only integration variables. Then the diagram of fig.3 is proportional
to the product $\Phi_+(R,q)\Phi_-(R,q)$ of the two semiclassical wave 
functions, each of which is represented by the functional integral:
$$
\Phi_{\pm}(R,q)=\int DN(\tau) DR(\tau) D\theta(\tau) Dp(\tau)
$$
\begin{equation}
\times \exp \left(-\frac{1}{\kappa} S_{gr}(R(\cdot),N(\cdot))-
\frac{1}{\kappa} \int_{-\infty}^{\tau_f} d\tau \left(
\frac{p^2N}{4\pi^2R^3}-i\sqrt{\kappa}\theta\dot{\pi}\right)\right),
\label{7*}
\end{equation}
where $S_{gr}$ is the gravitational action for metrics of the form
(\ref{2*}), the  boundary conditions are
\begin{equation}\label{8*}
p(\tau_f)=q, 
\end{equation}
(we consider the wormholes of global charge $q$),
\begin{equation}\label{8**}
R(\tau_f)=R, \qquad R(-\infty)=+\infty.
\end{equation}
The functional integral (\ref{7*}) is to be evaluated in the semiclassical
approximation; $\Phi_{-}$ and $\Phi_{+}$ contain the contributions of the 
classical solutions $R_-(\cdot)$ and $R_+(\cdot)$ (obeying boundary
conditions (\ref{8*}), (\ref{8**})) that do and do not pass through the
turning point ($\dot{R}=0,\dot{\theta}=0$), respectively.

Integrating eq.(\ref{7*}) over $\theta$, one obtains the factor 
$\delta(\dot{p})$. This means that $p=const=q$, and the integration over $p$
is immediate. The remaining integral is
$$
\Phi=\int DR DN e^{-S/\kappa},
$$
with
$$
S=\int_{-\infty}^{\tau_f} d\tau \left[ -6\pi^2NR \left(1
-\frac{R}{N}\frac{d}{d\tau}\left(\frac{\dot{R}}{N}\right)-
\frac{\dot{R}^2}{N^2}\right) + \frac{q^2N}{4\pi^2R^3}\right] -
\frac{6\pi^2}{N} R^2(\tau_f) \dot{R}(\tau_f).
$$
where we have taken into account the boundary term in the gravitational 
action. To evaluate the contributions of the two saddle-point configurations,
one first integrates over $R$ by
the Maslov technique \cite{Maslov} and then evaluates the integral over
$N$ by using the following definition of measure,
which coincides with the definition of ref.\cite{Mottola},
$$
\int DN F(\int N d\tau) \equiv \int dT F(T)
$$
The result is
\begin{equation}
\Phi_{\pm} (R,q)= \frac{1}{\sqrt{-\dot{R}_{\pm}(\tau_f)}} 
e^{-S[R_{\pm}(\cdot)]/\kappa}.
\label{9*}\end{equation}
Since $\dot{R}_+$ and $\dot{R}_-$ have opposite signs, the product
$$\Phi_+\Phi_-$$ is pure imaginary. 
This is the direct consequence of the fact that the original 
Giddings-Strominger solution (\ref{2**}) has a turning point. We conclude
that the amplitude of fig.1 is indeed purely imaginary.

{\bf 4.} The above argument is close in spirit to one of 
ref.\cite{ShvedovITEP}. Each half of the diagram of fig.3 may be viewed as
a contribution to the wave function of the ''ground state'' of the 
system ''parent universe + baby universes''. Indeed, one can argue
(cf. \cite{HartleHawking}) that the latter can be expressed through the
Euclidean functional integral over 4-manifolds with the following boundary
conditions: final 3-geometry is the argument of the wave function, while
initial 3-geometry is the classical ground state (flat space in our case,
not a point as in ref.\cite{HartleHawking}). The semiclassical wave function
of the system has therefore the form
$$
\Phi(R,q)=\Phi_+(R,q)+\Phi_-(R,q).
$$
Then one can construct the conserved current in minisuperspace,
$$
j_R=-i(\Phi\partial_R\Phi^* - \Phi^* \partial_R \Phi)
$$
that measures the flux of baby universes towards the singularity. From 
eq.(\ref{9*}) it follows that 
\begin{equation}\label{11*}
j_R=e^{-S_{GS}/\kappa},
\end{equation}
where $S_{GS}$ is the euclidean action of the wormhole of fig.1. The fact that
$j_R$ is non-zero is interpreted as a signal that the flat space is
unstable with respect to the emission of baby universes that then evolve
in their intrinsic time towards the singularity.

{\bf 5.} The latter argument suggests also that the picture of unstable flat
space is valid beyond the minisuperspace approximation. Indeed, in the
general case, there still exists a current that is conserved in superspace
\cite{DeWitt}. Its semiclassical value 
integrated over the surface of fixed $R$ in superspace
is still expected to be given 
by eq.(\ref{11*}), since there is non-zero flux of baby universes towards 
$R=0$ if the ''radiation'' boundary conditions are imposed.
Therefore, one expects that the amplitude of fig.1 will still be
pure imaginary when $O(4)$ -- asymmetric fluctuations about the wormhole
are properly accounted for. This expectation has to be confirmed by direct
analysis of general fluctuations about the wormhole solution.

{\bf 6.} The above arguments support the picture of baby universes that 
are steadily emitted by the parent universe and then evolve into the 
singularity (or towards $R=\infty$ in other models \cite{LRT,RT}).
This picture is similar to one emerging in (1+1)-dimensional
stringy model of parent and baby universes \cite{VR,VRNirov}.
In these cases one naturally expects that quantum coherence in the parent
universe is lost. However, the latter property has never been shown 
explicitly. Here we present a simple model which shares many features
of the above systems; the simplicity of the model enables one to see the
loss of quantum coherence explicitly and even obtain the equation for the
effective density matrix in the parent ''universe''.

Let us consider a quantum mechanical system (an analog of the ''parent
universe'') which is able to emit particles (''baby universes'').
The radiated particles can move along 
one-dimensional line in one direction (to the right) only. We choose the
Hamiltonian of each particle to be equal to $-i\partial/\partial x$, so
that all emitted particles  move with the velocity equal to 1.
Suppose that the particles can be emitted only at $x=0$, so that the full
Hamiltonian of the system under consideration is
\begin{equation}
H=T+\int dx a^+(x) (-i\frac{\partial}{\partial x}) a^-(x) + Va^+(0) +
V^+a^-(0).
\label{A1}
\end{equation}
Here $a^{\pm}(x)$ are creation and annihilation operators of ''baby 
universes'', operator $T$ corresponds to free evolution of the ''parent 
universe'', while $V$ is the operator in the ''parent universe'' state 
space which corresponds to emission of ''baby universes''.

Our purpose is to find whether there is loss of quantum coherence. The 
answer to this question depends on the boundary conditions imposed on the
radiated particles. If at the initial moment of time there existed ''baby
universes'' at $x=x_0<0$, they would move to the right, $x(t)=x_0+t$,
and at time $t=|x_0|$ they would reach the point $x=0$ and would
be able to interact with the ''parent universe'' according to eq.(\ref{A1}).
We see that the information about the ''baby universes'' residing at $x<0$
is not lost.

For example, if $V=V^+$ then there exist so-called ''$\alpha$-states''
\cite{Coleman}
$$
(a^+(0)+a^-(0))\Psi=\alpha\Psi;
$$
$$
\int dx a^+(x) (-i\frac{\partial}{\partial x}) a^-(x) \Psi =0.
$$
If the state of the full system is an $\alpha$-state at the initial  
moment of time,
 then the same property is satisfied at arbitrary time.
The effective evolution of the ''parent universe'' 
is described then by the Hamiltonian 
 $T+\alpha V$, where the ''constant of nature'' $\alpha$ is not specified
 by the theory but depends on the state of ''baby universes''.
 We see that  Coleman's ''$\alpha$-state'' interpretation of the
 Hamiltonian (\ref{A1}) is certainly possible. 

 However, one can consider another boundary condition which leads to
the loss of coherence for the ''parent universe''. 
 This condition is that there are no
 ''baby universes'' at $x<0$,
 \begin{equation}
 a^-(x)\Psi=0, x<0.
\label{A2}
 \end{equation}
 This condition is invariant under time evolution, since ''baby universes''
 are able to move only to the right and are created only at $x=0$. 
 If the condition (\ref{A2}) is imposed, the information 
 about ''baby universes'' is lost completely, as opposed to the ''$\alpha$-
 state'' case. Therefore, the initial reduced density matrix $\rho_0$
 of the ''parent universe'' determines in a unique fashion its effective
 density matrix $\rho$ at arbitrary time.                      
 We show in Appendix that the evolution equation for $\rho$ is a 
 special case of the equation of refs.\cite{BPS,EHNC}:
\begin{equation}
\dot{\rho}=-i[T,\rho] -\frac{1}{2}V^+V\rho -\frac{1}{2}\rho V^+V + V\rho V^+.
\label{AA1}
\end{equation}
It is straightforward to see \cite{BPS} that $Tr \rho^2$ decreases with
time, so that quantum coherence is indeed lost for the ''parent universe''
subsystem.

The authors are indebted to T. Banks, K. Lee, Kh. Nirov and P. Tinyakov for
helpful discussions. This work is supported in part by the 
Russian Foundation for Basic Research, project 96-02-17449a.

\section*{Appendix}

The purpose of this appendix is to obtain the evolution equation for the
reduced density matrix of the ''parent universe'', eq.(\ref{AA1}).
Consider the interaction representation
\begin{equation}
\Phi(t)=e^{iH_0t}\Psi(t)
\label{A2*}
\end{equation}
of the state vector $\Psi(t)=e^{-iHt}\Psi_0$; here
$$
H_0=T+\int dx a^+(x)(-i\frac{\partial}{\partial x}) a^-(x).
$$
Since 
$$
e^{iH_0t}a^{\pm}(\xi)e^{-iH_0t}=a^{\pm}(\xi-t),
$$
the vector $\Phi(t)$ satisfies the following equation
\begin{equation}
i\dot{\Phi}(t)=H_{int}(t)\Phi(t),
\label{A3}
\end{equation}
where
$$
V(t)=e^{iTt}Ve^{-iTt},
$$
$$
H_{int}(t)=V(t)a^+(-t) + V^+(t)a^-(-t).
$$
Let us show that under conditions (\ref{A2}) the operator $H_{int}(t)$ can be
replaced by the following effective Hamiltonian,
\begin{equation}
H_{eff}(t)=V(t)a^+(-t)-\frac{i}{2} V^+(t)V(t)
\label{A4}
\end{equation}
Note that the second term of eq.(\ref{A4}) is a direct analog of the 
imaginary part of the forward ''flat space $\to$ flat space'' amplitude 
discussed in this talk. 

To check  the formula (\ref{A4}), let us show that the following vector
$$
Texp(-i\int_0^t H_{eff}(t)dt)\Phi_0
$$
satisfies eq.(\ref{A3}) if the condition (\ref{A2}) is imposed on the 
initial state vector $\Phi_0$.
The left hand side of eq.(\ref{A3}) takes the form
\begin{equation}
H_{eff}(t)\Phi(t),
\label{A4*}
\end{equation}
while the right-hand side of eq.(\ref{A3}) is as follows:
$$
\sum_{n=0}^{\infty} 
(V(t)a^+(-t)+V^+(t)a^-(-t))
\int_{t>t_1>...>t_n>0} dt_1...dt_n (-i)^n\times
$$
\begin{equation}
\times(V(t_1)a^+(-t_1)-\frac{i}{2} V^+(t_1)V(t_1))
...
(V(t_n)a^+(-t_n)-\frac{i}{2} V^+(t_n)V(t_n))
\Phi_0.
\label{A5}
\end{equation}
Because of conditions (\ref{A2}), the 
terms with $[a^-(-t),a^+(-t_i)]$ give 
$\delta$-functions $\delta(t-t_i)$. If $i\ge 2$ then the range of 
$t_1$ collapses to a point, so that the contributions of terms containing
$a^-(-t)a^+(t_i)$, $i\ge 2$ vanish. The only non-vanishing contribution
is given by the term $[a^-(-t),a^+(-t_1)]$. We now have to define
the integrals like
\begin{equation}
\int_{t_2<t_1\le t} \delta(t_1-t) V^+(t) V(t_1) dt_1.
\label{A6}
\end{equation}
To resolve this ambiguity, let us regularize the Hamiltonian (\ref{A1}) by
the replacement
$$
a^+(0) \to \int a^+(\xi) f(\xi) d\xi,
$$
where $f$ is a sharply peaked real function. After this regularization the 
quantity
$$
[a^-(-t),a^+(-t_1)]=\delta(t-t_1)
$$
is replaced by
$$
\int d\xi d\xi_1 [a^-(-t+\xi),a^+(-t_1+\xi_1)] f(\xi)f(\xi_1)=
\int d\xi f(\xi) f(t_1-t+\xi).
$$
The integral (\ref{A6}) takes the form $\frac{1}{2} V^+(t)V(t)$, so that
the quantity (\ref{A5}) coincides with (\ref{A4*}).

Making use of the effective Hamiltonian (\ref{A4}), it is 
straightforward to obtain 
eq.(\ref{AA1}).
 Namely, consider the trace with respect to ''baby universes'' 
of the matrix 
$|\Phi(t)><\Phi(t)|$
$$
\rho_{\Phi}(t)=Tr_{baby} 
|\Phi(t)><\Phi(t)|.
$$
Since 
$$
Tr_{baby} a^+(-t)
|\Phi(t)><\Phi(t)|=0,
$$
and
$$
Tr_{baby} a^+(-t_1)
|\Phi(t_1)><\Phi(t_2)|
a^-(-t_2)=\delta(t_1-t_2) \rho_{\Phi}(t_1),
$$
the difference between matrices $\rho_{\Phi}$ at times $T+t$ and $T$
is
$$
\rho_{\Phi}(T+t)-\rho_{\Phi}(T)=t [
-\frac{1}{2} V^+(T)V(T)\rho_{\Phi}(T)
-\frac{1}{2} \rho_{\Phi}(T)V^+(T)V(T)+ 
$$
$$
+
V(T)\rho_{\Phi}(T)V^+(T) ]
+O(t^2).
$$
Making use of 
 formula (\ref{A2*}) one obtains the equation for the reduced density
matrix, eq. (\ref{AA1}).

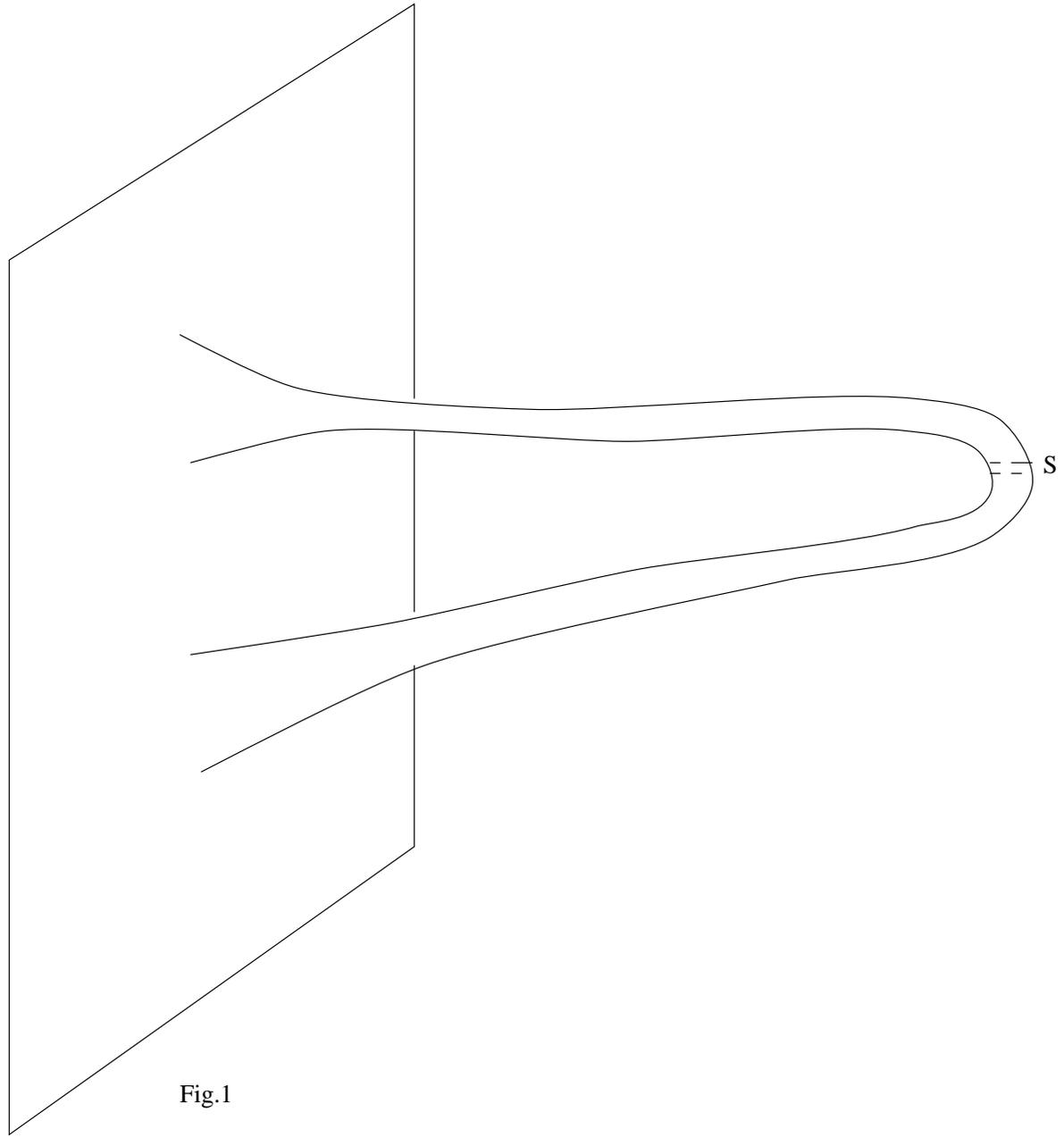
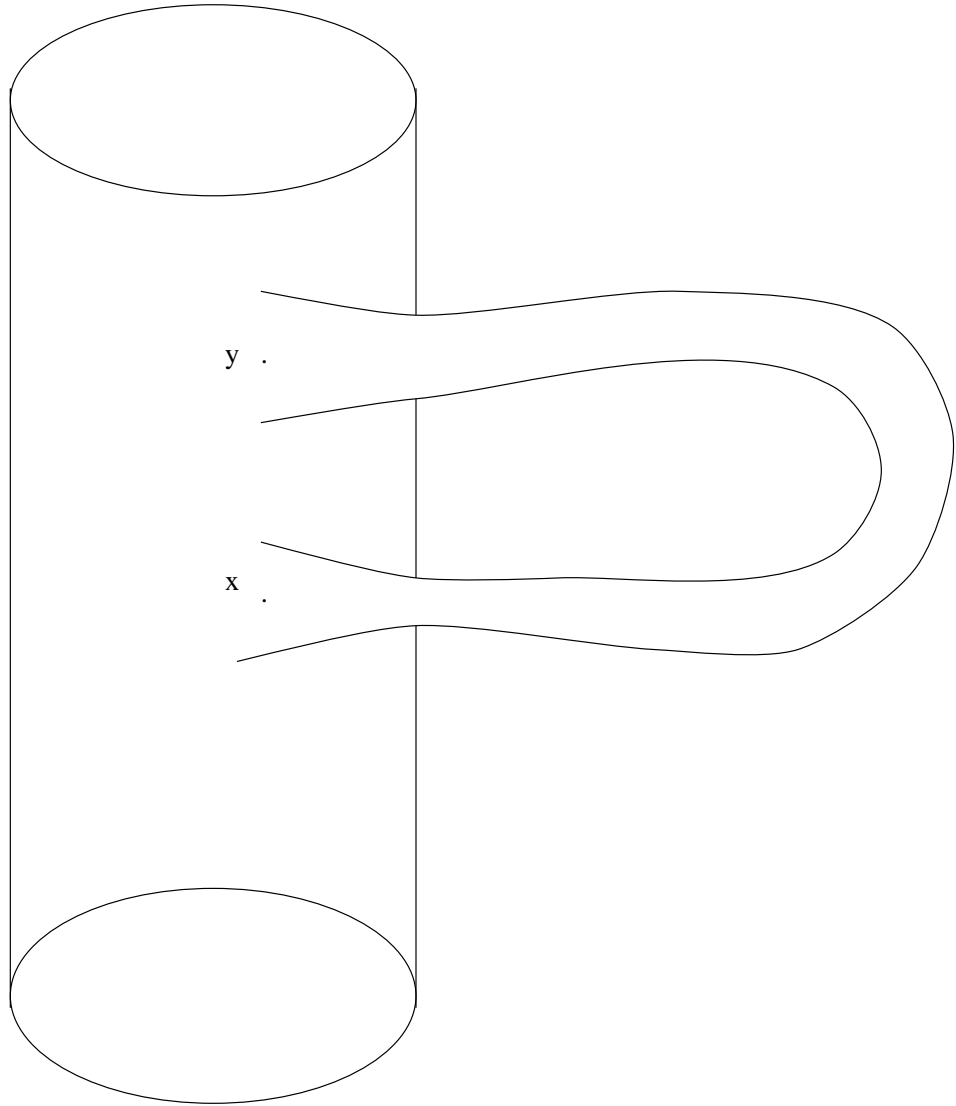
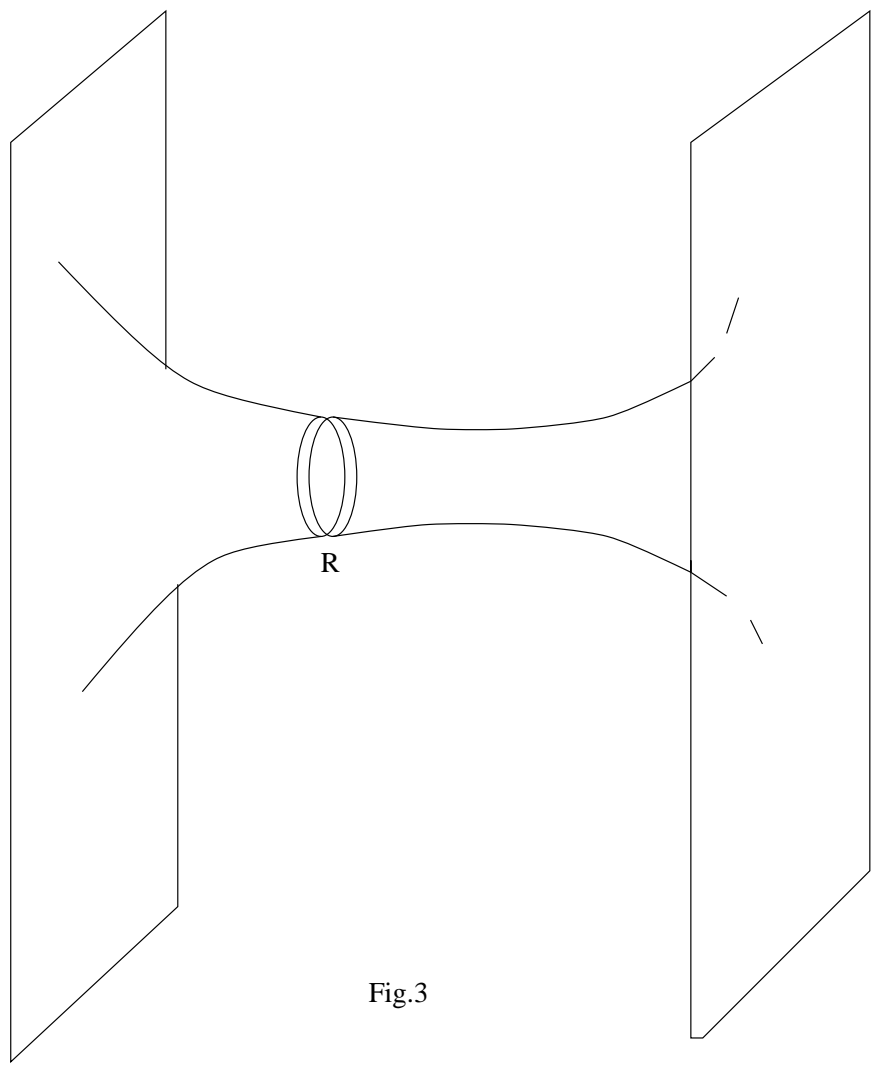
\end{document}